\documentclass[useAMS,usenatbib,twocolumn]{mn2e}
\usepackage{subfigure}
\usepackage{rotating}
\usepackage{times}
\usepackage{graphicx}
\usepackage{lscape}

\def\etal{{\it et al.}}

\def\sqiggt{\hbox{\rlap{\lower.55ex \hbox {$\sim$}}\kern-.05em \raise.4ex \hbox{$>$}\,}}
\def\sqiglt{\hbox{\rlap{\lower.55ex \hbox {$\sim$}}\kern-.05em \raise.4ex \hbox{$<$}\,}}

\title[{The unusual X-ray emission of GRB 090515}]{The unusual X-ray emission of the short {\it Swift} GRB 090515: Evidence for the formation of a magnetar?}
\author[A. Rowlinson \etal ]{A. Rowlinson$^{1}$\thanks{E-mail:
bar7@star.le.ac.uk}, P.~T. O'Brien$^{1}$, N.~R. Tanvir$^{1}$, B. Zhang$^{2}$, P.~A. Evans$^{1}$, N. Lyons$^{1}$,\and 
A.~J. Levan$^{3}$, R. Willingale$^{1}$, K.~L. Page$^{1}$, O. Onal$^{4}$, D.~N. Burrows$^{5}$, A.~P. Beardmore$^{1}$, \and 
T.~N. Ukwatta$^{6,7}$, E. Berger$^{8}$, J. Hjorth$^{9}$, A.~S. Fruchter$^{10}$, R.~L. Tunnicliffe$^{3}$, \and 
D.~B. Fox$^{5}$, A. Cucchiara$^{5}$ \\$^{1}$Department of Physics \& Astronomy,University of Leicester, University Road, Leicester, LE1 7RH, UK\\$^{2}$Department of Physics, University of Nevada, 4505 South Maryland Parkway, Las Vegas, NV 89154, USA\\$^{3}$Department of Physics, University of Warwick, Coventry CV4 7AL \\$^{4}$Istanbul University, Faculty of Science, Department of Astronomy and Space Sciences, 34119 University, Istanbul, Turkey \\$^{5}$ Department of Astronomy and Astrophysics, Pennsylvania State University, 525 Davey Lab, University Park, PA 16802, USA \\$^{6}$ The George Washington University, Washington, D.C. 20052, USA \\$^{7}$ NASA Goddard Space Flight Centre, Greenbelt, MD 20771, USA \\$^{8}$ Harvard-Smithsonian Center for Astrophysics, 60 Garden Street, Cambridge, MA 02138, USA\\$^{9}$ Dark Cosmology Centre, Niels Bohr Institute, University of Copenhagen, Juliane Maries Vej 30, 2100 Copenhagen, Denmark \\$^{10}$ Space Telescope Science Institute, 3700 San Martin Drive, Baltimore, MD 21218, USA}

\begin{document}

\date{Accepted 00. Received 00; in original form 00}

\pagerange{\pageref{firstpage}--\pageref{lastpage}} \pubyear{000}
\maketitle            

\label{firstpage}

\begin{abstract}
  
The majority of short gamma-ray bursts (SGRBs) are thought to originate from 
the merger of compact binary systems collapsing directly to form a black hole. 
However, it has been proposed that both SGRBs and long gamma-ray bursts (LGRBs) 
may, on rare occasions, form an unstable millisecond pulsar (magnetar) prior to 
final collapse. GRB 090515, detected by the {\it Swift} satellite was extremely 
short, with a $T_{90}$ of 0.036$\pm$0.016 s, and had a very low fluence of 
$2 \times 10^{-8}$ erg cm$^{-2}$ and faint optical afterglow. Despite this, 
the 0.3 -- 10 keV flux in the first 200 s was the highest observed for a SGRB 
by the {\it Swift} X-ray Telescope (XRT). The X-ray light curve showed an 
unusual plateau and steep decay, becoming undetectable after $\sim$500 s. 
This behaviour is similar to that observed in some long bursts proposed to 
have magnetars contributing to their emission.

In this paper, we present the {\it Swift} observations of GRB 090515 and compare 
it to other gamma-ray bursts (GRBs) in the {\it Swift} sample. Additionally, we 
present optical observations from Gemini, which detected an afterglow of 
magnitude $26.4\pm0.1$ at T+ 1.7 hours after the burst. We discuss potential causes
of the unusual 0.3 -- 10 keV emission and suggest it might be energy injection 
from an unstable millisecond pulsar. Using the duration and flux of the plateau of 
GRB 090515, we place constraints on the millisecond pulsar spin period and magnetic 
field.

\end{abstract}


\section{Introduction}

Thirty years after the discovery of Gamma-ray Bursts (GRBs) by the {\it Vela} 
satellites \citep{klebesadel1973}, the first X-ray afterglow was detected for 
GRB 970228 by the {\it Beppo-SAX} satellite \citep{costa1997}. With the 
increased accuracy for the position provided by X-ray afterglows, it was 
possible to identify the optical afterglow and the host galaxies of many GRBs.

With the launch of the {\it Swift} satellite \citep{gehrels2004}, the X-ray 
afterglow has been studied in great detail, placing tighter constraints on 
models for GRB emission. Additionally, {\it Swift} has enabled the detection of Short 
GRB (SGRB) X-ray afterglows, allowing them to be 
directly compared to Long GRB (LGRB) afterglows \citep{gehrels2005, barthelmy2005, 
hjorth2005}. \cite{nousek2006} and \cite{zhang2006} described the 
``canonical GRB light curve'' as three stages comprising a steep decline followed 
by a shallower decay and then a final decay phase. \cite{obrien2006} showed that not 
all X-ray light curves for GRBs are of the ``cannonical'' variety. They and 
\cite{willingale2007} suggested that the X-ray light curve comprises two main 
components, the prompt emission and the afterglow. The relative strength of these 
components determines the observed X-ray light curve. A more recent study of all
{\it Swift} X-ray afterglows by \cite{evans2008} has reinforced these findings. The 
initial steep decay following the prompt emission typically has a power law 
decay with index $\alpha \sim 2-5$, where $f \propto t^{-\alpha}$ (t is the time after 
the burst in seconds and f is the flux) \citep{obrien2006}. 

Multi-wavelength observations have associated LGRBs with type Ibc core 
collapse supernovae at cosmological distances \citep[e.g.][]{hjorth2003, 
stanek2003}, although not all such supernovae produce long GRBs \citep{woosley2006}.
The progenitors of SGRBs are less well understood, but the most popular theory is that 
they originate from the merger of compact binary systems, for example 
neutron stars or a neutron star and a black hole \citep{lattimer1976,eichler1989,
narayan1992}. It has also been suggested 
that both LGRB and SGRB progenitors could produce an unstable millisecond pulsar. This is 
expected to contribute a small fraction of the GRB population \citep{usov1992, duncan1992, 
dai1998a, dai1998b, zhang2001}. \cite{troja2007} and \cite{lyons2009} found examples of LGRBs 
that have an observable plateau and steep decay in the X-ray light curve, which have been 
intepreted as caused by energy injection by an unstable millisecond pulsar which then 
collapses. Magnetar models have also been proposed to explain late central engine activity 
in SGRBs, for example late time plateaus in the X-ray afterglows \citep{fan2006, yu2010, 
dallosso2010} and X-ray flares \citep{fan2005,gao2006}. Here we present an analysis of GRB 
090515 which is the best case for an early X-ray plateau in an SGRB.

GRB 090515 was one of the shortest GRBs observed by {\it Swift}, with among the lowest 
fluence, yet for $\sim$200 s it had the brightest SGRB X-ray afterglow and did not appear 
to be fading until a sudden steep decline at $\sim$ 240 s. After the first orbit, it was not 
detected again. Explaining this unusual X-ray behaviour is the subject of this paper. We 
describe the observations of GRB 090515 in section 2, compare it to other GRBs in section 
3 and discuss the potential origin of the unusual X-ray emission in section 4. Throughout 
the paper we adopt a cosmology with $H_0 =  71$ km\,s$^{-1}$\,Mpc$^{-1}$, $\Omega_m = 0.27$, 
$\Omega_\Lambda = 0.73$. Errors are quoted at 90\% confidence for X-ray data and at 1$\sigma$ 
for optical data.

\section{Observations}

\subsection{Swift Observations}
\begin{figure}
\centering
\includegraphics[width=8.2cm]{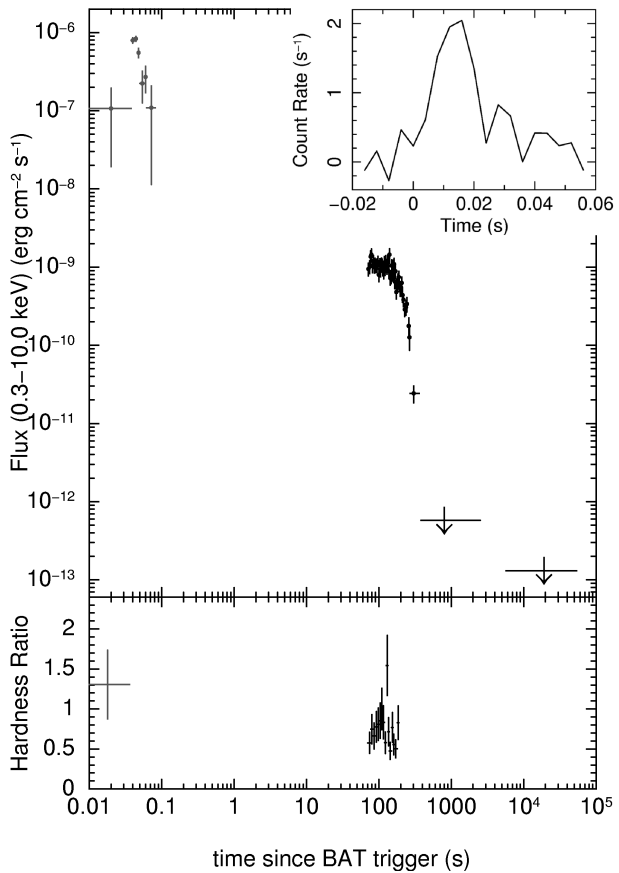}
\caption{The combined light curve for GRB 090515, in grey are the BAT data and black are the XRT data. 
In the lower box
there is the hardness ratio for the BAT data ((50 -- 100) keV/(25 -- 50) keV) in grey and 
the hardness ratio for the XRT data ((1.5 -- 10) keV/(0.3 -- 1.5) keV) in black. 
Inset is the BAT count rate per detector light curve with linear time.}
\label{fig1}
\end{figure}

All analysis has been performed by using standard routines in {\sc heasoft}, 
{\sc xspec}, {\sc qdp} and the automatic X-ray Telescope \citep[XRT,][]{burrows2005} data 
products produced by the UK Swift Science Data Centre \citep{evans2007, evans2008}.

{\it Swift} triggered on GRB 090515 at 04:45:09 UT on 15th May 2009 with BAT position
RA = 10h 56m 41s and Dec = +14$^{\circ}$ 27$^{\prime}$ 22$^{\prime\prime}$ 
\citep{beardmore2009}. The Ultra-Violet and Optical Telescope (UVOT) enhanced 
refined XRT position was RA = 10h 56m 36.11s and Dec = +14$^{\circ}$ 26$^{\prime}$ 
30.3$^{\prime\prime}$ with an uncertainty of 2.7$^{\prime\prime}$ \citep{osbourne2009}.

The $T_{90}$ duration of GRB 090515 was $0.036 \pm 0.016$ s \citep{barthelmy2009}. The 
spectrum of the prompt gamma-ray emission can be fit by a single power law,
of photon index  $\Gamma_{\gamma} = 1.6 \pm 0.2$ \citep{barthelmy2009}. The fluence 
is $2.0 \pm 0.8 \times 10^{-8}$ erg cm $^{-2}$ and the peak 
photon flux is $5.7 \pm 0.9$ ph cm$^{-2}$ s$^{-1}$. All values are in the 15 -- 150 keV energy 
band. The BAT light curve is shown in Figure \ref{fig1} as the grey data points and also 
shown in the inset with linear time. The BAT count rates were converted to flux in the 
energy band 0.3 -- 10 keV using the average spectral index for the BAT and the XRT spectra. 
There is no evidence of extended emission detected in the BAT energy range \citep{norris2010}.

We completed a spectral lag analysis for GRB 090515 using the cross correlation function 
method described in \cite{ukwatta2010}, the 8 ms time binned lightcurve and BAT channels 
1, 2 and 3. Channel 4 did not detect enough emission to make a lag measurement. The lag 
times are (with 1$\sigma$ errors): lag(Ch2-Ch1)$=6\pm4$ ms, lag(Ch3-Ch2)$=3\pm2$ ms and 
lag(Ch2-Ch1)$=10\pm4$ ms. Typically SGRBs have negligble lag times \citep{norris2006,yi2006} 
and LGRBs have typical lag times ranging from 20 ms to $\sim$1000 ms \citep{ukwatta2010}, so 
it is interesting that GRB 090515 appears to have a small lag time.

The X-ray spectrum in the 0.3 -- 10 keV energy band is best fit by an absorbed
power law with $\Gamma_{X} = 1.88 \pm 0.14$ and $N_{\rm H}=6.1^{+3.0}_{-2.8} \times 
10^{20}$ cm$^{-2}$, in excess of the Galactic $N_{\rm H}=1.9 \times 10^{20}$ cm$^{-2}$ 
\citep{beardmore2009b}. The X-ray light 
curve is best fit by a broken power law with 2 breaks giving a reduced $\chi^{2}_{\nu}$ 
of 0.86. The initial decay is relatively flat 
($\alpha_{1}=0.29^{+0.08}_{-0.27}$) with a break at 
$T_{1} = 156.2^{+9.3}_{-26.2}$s followed by a steeper decay of 
$\alpha_{2}=2.51^{+0.38}_{-0.70}$. At $T_{2} = 240.8^{+7.4}_{-9.8}$s it
breaks to an extremely steep decay of 
$\alpha_{3} > 9$. Although, we have fitted the X-ray light curve using a broken power law, we 
note that the decay appears to be a smooth curve. The 
X-ray light curve is shown in Figure \ref{fig1} as the black data points and the lower panel 
shows the hardness ratio for the gamma-ray emission (in grey), 
i.e. the ratio of the 50 -- 100 keV emission to the 25 -- 50 keV emission, and
the hardness ratio of the X-ray emission (in black, (1.5 -- 10) keV/(0.3 -- 1.5) keV). 
The hardness ratio is fairly constant during the plateau, with the exception
of a point at $\sim$120 s that could be a flare and does correspond to a small 
peak in the X-ray light curve, but this may just be noise. There are 
insufficient counts to characterise the hardness ratio during the decay.

\subsection{Early Optical Observations}

\begin{table*}
\begin{center}
\caption{The optical observations of the field of GRB 090515.}
\begin{tabular}{| c | c | c | c | c | c |}
\hline
Telescope & Mid point time after trigger & Exposure Time & Band & Upper Limit  & 
Flux Upper Limit\\
   & (s) & & & (magnitude) &  (erg~cm$^{-2}$s$^{-1}$Hz$^{-1}$) \\
\hline
KAIT & 20 & 540 & R & 19.1$^{(1)}$ &  $6.6 \times 10^{-28}$ \\
Super LOTIS & 43 & 10 & R & 17.7$^{(2)}$ & $2.4 \times 10^{-27}$ \\
ROTSE III & 86 & 67 & R & 18.4$^{(3)}$ & $1.3 \times 10^{-27}$ \\
UVOT & 142 & 146 & White & 20.35$^{(4)}$ & $2.1 \times 10^{-28}$ \\
UVOT & 1228 & 488 & White & 21.24$^{(4)}$ & $9.2 \times 10^{-29}$ \\
KAIT & 2078 & 540 & R & 20.5$^{(1)}$ & $1.8 \times 10^{-28}$ \\
Lick & 2286 & 60 & R & 21.3$^{(5)}$ & $8.7 \times 10^{-29}$ \\
ROVOR & 5496 & 4200 & R & 21.4$^{(6)}$ & $7.9 \times 10^{-29}$ \\

\hline
\end{tabular}
\end{center}
$^{(1)}$ \cite{li2009}, 
$^{(2)}$ \cite{williams2009}, 
$^{(3)}$ \cite{rujopakarn2009}, 
$^{(4)}$ \cite{seigel2009},\\ 
$^{(5)}$ \cite{perley2009}, 
$^{(6)}$ \cite{pace2009}. 
\end{table*}

The field of GRB 090515 was observed at early times by several optical telescopes but none detected an 
optical afterglow. The upper limits of the R band and white filter observations 
are given in Table 1. During the plateau phase, we can predict the optical 
flux density, assuming that the X-ray and optical emission are from the same emitting region. 
If there is not a cooling break in the spectrum (i.e. $\Gamma_{X} = \Gamma_{OX}$) then 
we would expect the optical flux to be  $1.7 \times 10^{-26}$ erg cm$^{-2}$ s$^{-1}$ Hz$^{-1}$, 
corresponding to an apparent magnitude of $R=15.6$. This is brighter than all of the optical upper 
limits during the plateau, so we should have 
observed the optical afterglow. However, if there were a cooling break in the spectrum 
between optical and X-ray then  $\Gamma_{OX} = \Gamma_{X} - 0.5$ and, in this case, the 
optical flux density would be
$8.7 \times 10^{-29}$ erg cm$^{-2}$ s$^{-1}$ keV$^{-1}$Hz$^{-1}$, corresponding to an apparent
magnitude of $R=21.3$. This is slightly deeper than the optical upper limit provided
by UVOT. Therefore, if the optical emission was from the same emitting region as the X-ray 
and there is a cooling break in the spectrum, there is a slim chance that the optical flux 
was below the observed limits so the non-detection is consistent with the X-ray data.

\subsection{Gemini Observations}

\begin{table*}
\caption{Log of Gemini observations. \label{table:log}}
\begin{tabular}{@{}ccccccccc@{}}
\hline
Epoch & Date start & Start time after trigger & Exposure time & Filter & Seeing & Airmass & Magnitude & Flux \\ 
& (UT) & (s) & (s) & & (arcsec) & & & (erg~cm$^{-2}$s$^{-1}$Hz$^{-1}$) \\
\hline
1 & May 15 06:27 UT & $\sim$ 6100  & 1800 & r & 0.5 & 1.021 & 26.36 $\pm$ 0.12 & $8.2 \times 10^{-31}$ \\
2 & May 16 05:44 UT & $\sim$ 9$\times$10$^{4}$ & 1800 & r & 1 & 1.005 & 26.54 $\pm$ 0.33 & $6.95 \times 10 ^{-31}$\\
3 & November 28 14:20 UT & $\sim$ 1.6$\times$10$^{7}$  & 2800 & r & 0.8 & 1.226 & $>$ 27.4 & $< 4.55 \times 10 ^{-31}$ \\
\hline
\end{tabular}
\end{table*}

We obtained optical observations of GRB 090515 using Gemini North and GMOS beginning at 06:26 
UT, approximately 1.7 hours after the burst, with a second epoch observation being taken on the 
subsequent night, and a final comparison epoch on 28 November 2009. The images were obtained in 
the $r$-band, and were reduced via the standard {\sc iraf} Gemini tasks \citep{tody1993}. The 
image conditions for our first epoch were excellent, with seeing of 0.5$^{\prime\prime}$, 
resulting in extremely deep imaging in our total exposure time of 1800s. A full log of 
observations is shown in Table~\ref{table:log}.

\begin{figure}
\centering
\includegraphics[width=8.2cm]{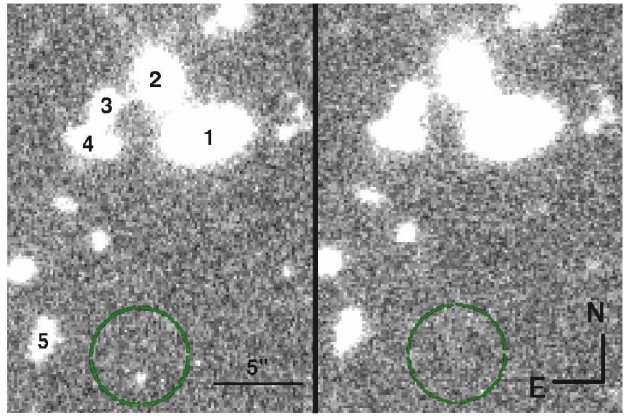}
\caption{The circle marks the location of the XRT enhanced position of GRB 090515 
on the Gemini observation from epoch 1 on the left and epoch 3 on the right. An 
optical afterglow candidate is observed within the error circle. Labeled are the 
brightest nearby galaxies.}
\label{fig1_opt}
\end{figure}

Within the refined XRT error circle we locate a single source at RA = 10h 56m 35.89s and Dec = 
+14${^\circ}$ 26$^{\prime}$ 30.0$^{\prime\prime}$, with a magnitude of $r=26.36 \pm 0.12$, 
calibrated against existing SDSS observations of the field, shown in Figure \ref{fig1_opt}. 
This source is still visible, but at lower significance in our shallower images obtained on 
16 May ($r = 26.54 \pm 0.33$). In our final epoch there is no source visible at the afterglow 
location, to a limiting magnitude of $r>27.4$ confirming a fading counterpart. We therefore 
conclude that this is the optical afterglow of GRB 090515. At $r=26.36$, this is the 
faintest GRB afterglow ever discovered at similar times after the burst, and confirms the 
necessity of rapid and deep observations with 8-metre class observatories. As the observed 
X-ray absorption is relatively low ($N_{\rm H} \sim 6 \times 10^{20}$ cm$^{-2}$), the faint 
optical afterglow is unlikely to be a consequence of extinction (unless it is at high redshift). 
The optical afterglow has a relatively flat lightcurve, with a decay slope of 
$0.06^{+0.32}_{-0.19}$.

Comparing this afterglow to the sample in \cite{nysewander2009}, we note 
that this is the first SGRB with a fluence below $10^{-7}$ erg cm$^{-2}$ with a detected 
optical afterglow. Additionally, the afterglow at 1.7 hours is fainter than all the observed 
optical afterglows at 11 hours. GRB 080503 also had an initially very faint optical afterglow, 
but it then rebrightened to a peak of $r\sim25.5$ at 1 day and no host galaxy was 
identified \citep{perley2009b}.

Assuming there is not a cooling break in the spectrum, i.e. $\Gamma_{X} = \Gamma_{OX}$, 
we predict that the X-ray flux, 0.3 - 10 keV, at the time of the optical observations should be 
$6.6 \times 10^{-15}$ erg cm$^{-2}$ s$^{-1}$. This is consistent with the observed upper 
limit. 

Labeled in Figure \ref{fig1_opt} are the five brightest nearby galaxies and Table 
\ref{table:galaxies} provides their magnitudes and offsets from the GRB location. These 
galaxies are candidates for the host galaxy of GRB 090515, with significant offsets, 
or the burst could be associated with a significantly fainter underlying host galaxy.

\begin{table}
\caption{Photometry of the nearby galaxies, as labeled in Figure \ref{fig1_opt}, 
calibrated using SDSS observations. \label{table:galaxies}}
\begin{center}
\begin{tabular}{@{}ccccccccc@{}}
\hline
Galaxy & Magnitude & Offset (") \\
\hline
1 & 20.2 $\pm$ 0.1 & 14 \\
2 & 21.3 $\pm$ 0.1 & 16 \\
3 & 22.5 $\pm$ 0.1 & 15 \\
4 & 22.6 $\pm$ 0.1 & 13 \\
5 & 23.4 $\pm$ 0.1 & 6 \\
\hline
\end{tabular}
\end{center}
\end{table}

\section{Comparison to other GRBs}

\begin{table}
\caption{The GRBs considered in detail in this paper. \label{table:grbs}}
\begin{tabular}{@{}ccccccccc@{}}
\hline
GRB & T$_{90}$ & $\Gamma$ (15 -- 150 keV) & Fluence (15 -- 150 keV) \\
 & (s) & & ($10^{-8}$ erg cm$^{-2}$) \\
\hline
090515 & 0.036$\pm$0.016 & 1.6$\pm$0.2 & 2.0 $\pm$ 0.8 $^{(1),(2)}$\\
\hline
090607 & 2.3$\pm$0.1 & 1.25$\pm$0.30 & 11$\pm$2 $^{(3)}$\\
080520A & 2.8$\pm$0.7 & 2.90$\pm$0.51 & 5.5$\pm$1.4 $^{(4)}$\\
080503 & 170$\pm$20 & 2.00$\pm$0.13 & 200$\pm$10 $^{(5)}$\\
070724A & 0.4$\pm$0.04 & 1.81$\pm$0.33 & 3.0$\pm$0.7 $^{(6)}$\\
070616 & 402$\pm$10 & 1.61$\pm$0.04 & 1920$\pm$30 $^{(7)}$\\
070209 & 0.10$\pm$0.02 & 1.55$\pm$0.39 & 1.1$\pm$0.3 $^{(8)}$\\
060717 &3.0$\pm$1 & 1.72$\pm$0.38 & 6.5$\pm$1.6 $^{(9)}$\\
051221B & 61$\pm$1 & 1.48$\pm$0.18 & 113$\pm$13 $^{(10)}$ \\
051105 & 0.028$\pm$0.004 & 1.33$\pm$0.35 & 2.0$\pm$0.46 $^{(11)}$\\
050813 & 0.6$\pm$0.1 & 1.19$\pm$0.33 & 4.4$\pm$1.1 $^{(12)}$\\
050509B & 0.048$\pm$0.022 & 1.5$\pm$0.4 & 0.78$\pm$0.22 $^{(13)}$\\
050421 & 10.3$\pm$2 & 1.7$\pm$0.4 & 8.8$\pm$2.9 $^{(14)}$\\
\hline
\end{tabular}
$^{(1)}$ \cite{barthelmy2009}
$^{(2)}$ \cite{sakamoto2009}
$^{(3)}$ \cite{barthelmy2009b}
$^{(4)}$ \cite{sakamoto2008} 
$^{(5)}$ \cite{ukwatta2008} 
$^{(6)}$ \cite{parsons2007}
$^{(7)}$ \cite{sato2007}
$^{(8)}$ \cite{sakamoto2007}
$^{(9)}$ \cite{markwardt2006}
$^{(10)}$ \cite{fenimore2005}
$^{(11)}$ \cite{barbier2005b}
$^{(12)}$ \cite{sato2005}
$^{(13)}$ \cite{barthelmy2005b}
$^{(14)}$ \cite{sakamoto2005}
\end{table}

\begin{figure}
\centering
\includegraphics[width=8.2cm]{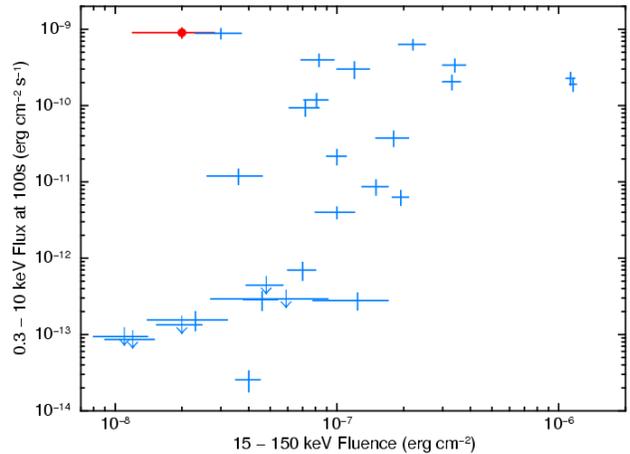}
\caption{The fluence in the energy band 0.3 -- 10 keV versus the 15 -- 150 keV 
flux for all Swift SGRBs which were observed at 100s after the trigger time. 
The filled circle marks the location of GRB 090515.}
\label{fig_fluence}
\end{figure}

\begin{figure}
\centering
\includegraphics[width=8.4cm]{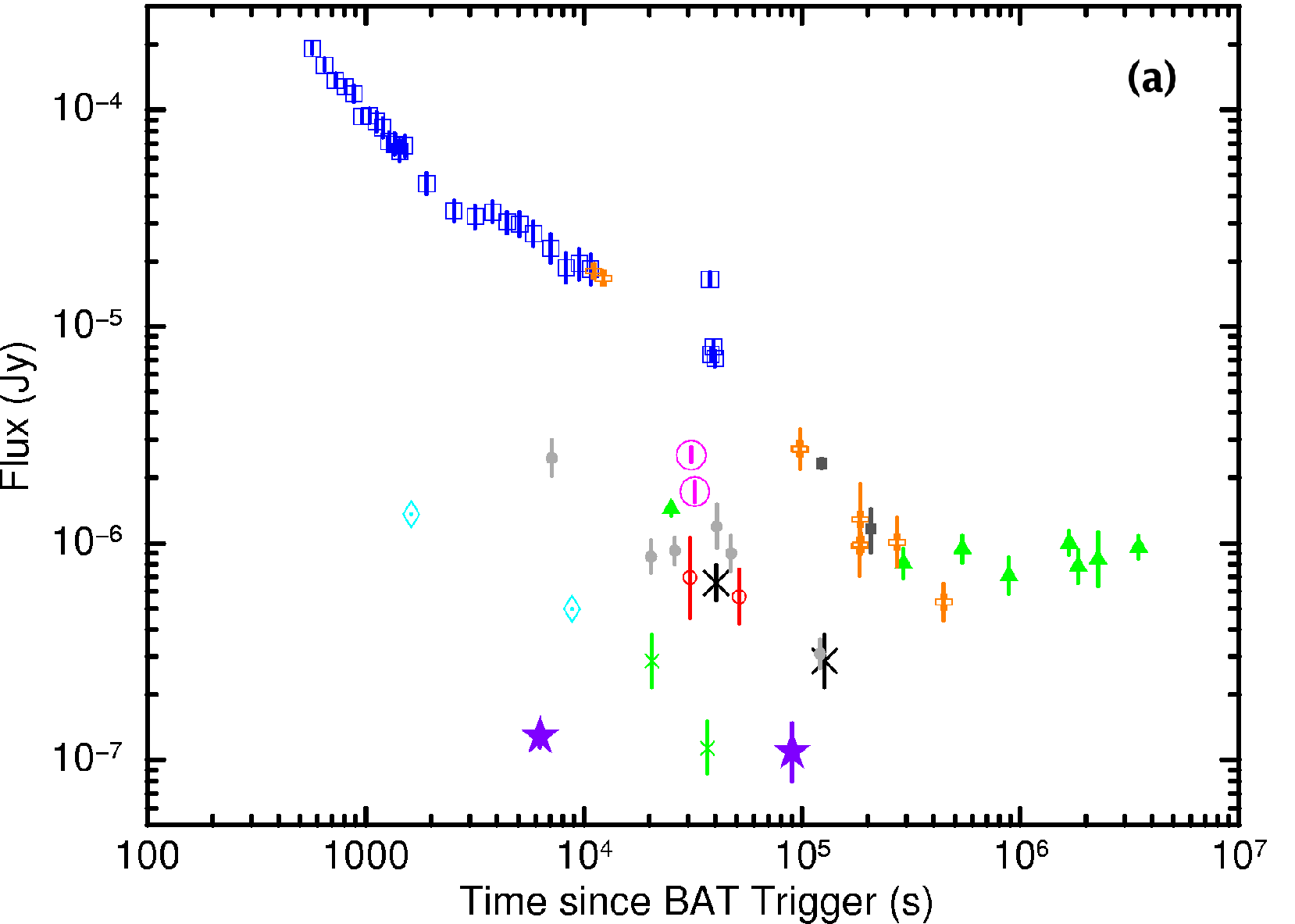}
\includegraphics[width=8.4cm]{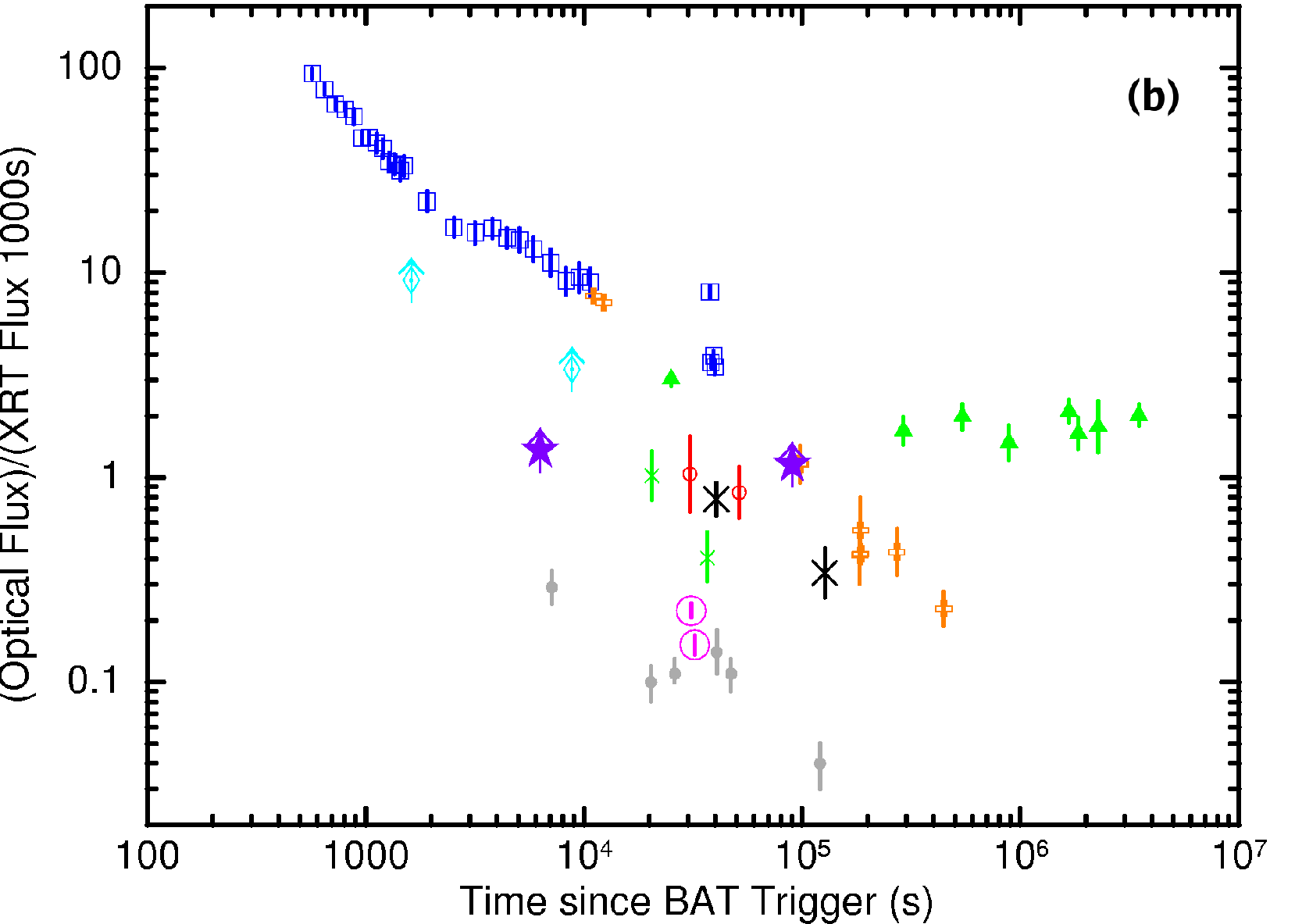}
\includegraphics[width=8.4cm]{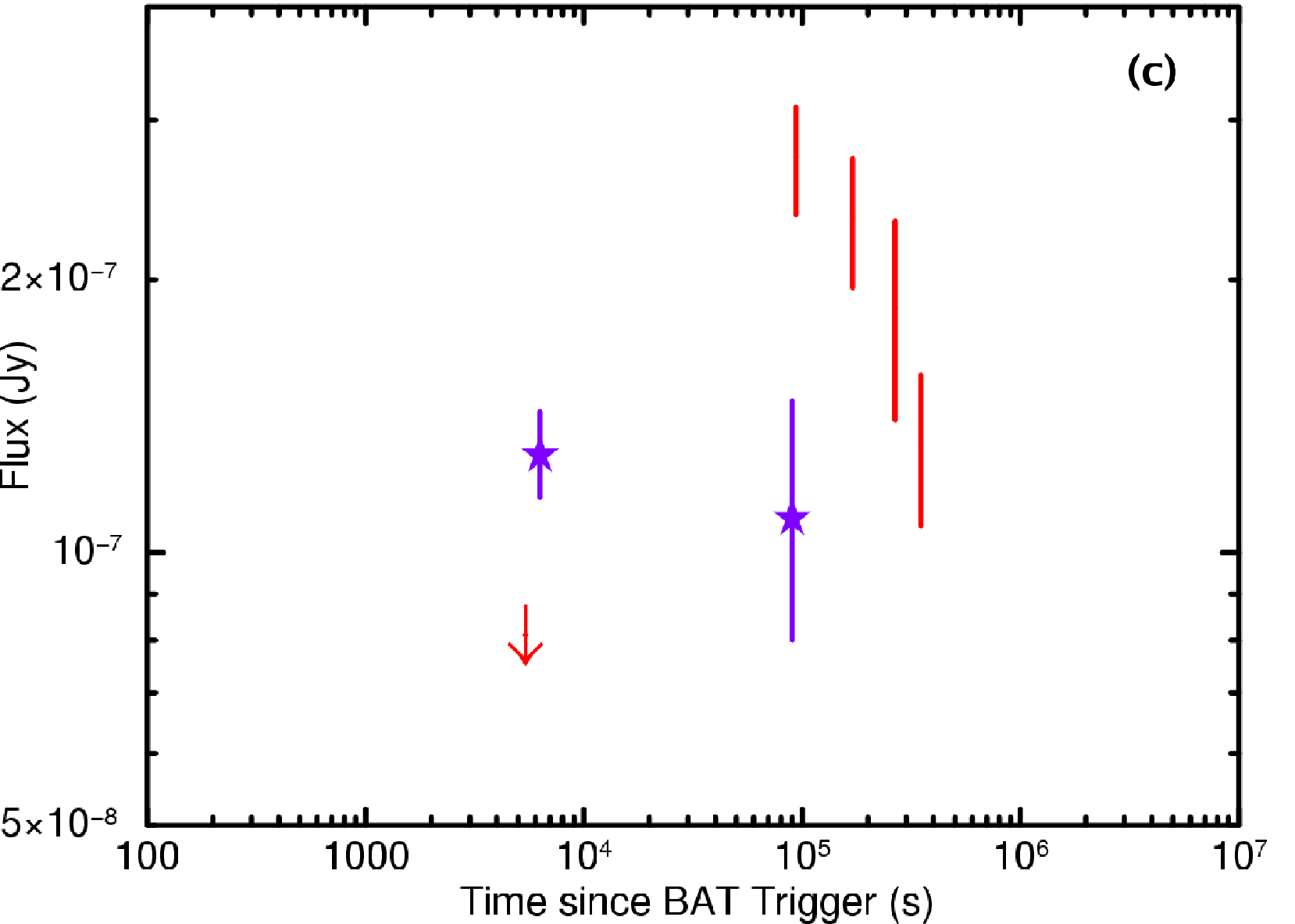}

\caption{(a) The optical flux light curves for all observed SGRB afterglows in the R band.
(b) Normalised using the XRT flux at 1000 s. Colour scheme: GRB 090515 - purple star, 
GRB 091109B - small green X \citep{levan2009,malesani2009}, GRB 090426 - dark blue open square
\citep{antonelli2009,xin2010}, GRB 090305 - light blue open diamond \citep{cenko2009,berger2009b}, 
GRB 080905A - small red open circle \citep{rowlinson2010}, GRB 071227 - green filled triangle \citep{berger2007}, 
GRB 070809 - large black X \citep{perley2007,perley2008}, GRB 061201 - large pink open circle \citep{stratta2007}, 
GRB 060121 - dark grey filled circle \citep{levan2006}, GRB 051221A - orange open cross \citep{soderberg2006}, 
GRB 050709 - light grey filled square \citep{hjorth2005}. (c) The optical flux light curve for GRB 
090515 (purple stars) with GRB 080503 (red).}
\label{optical_lcs}
\end{figure}

The XRT light curve of the low fluence GRB 090515 is unusual as it goes from being the 
brightest SGRB in X-rays to one of the faintest within seconds. The fluence 
in X-rays during the plateau is significantly higher than the fluence in 
gamma-rays. Additionally, the final decay is the steepest decay observed to date
\citep{evans2008}. The X-ray spectral index of GRB 090515 is not unusual compared to other 
SGRBs. In Table \ref{table:grbs}, we provide a summary of the properties of the long 
and short GRBs to which we compare GRB 090515 in detail. 

In Figure \ref{fig_fluence}, we show the 15 -- 150 keV fluence and 0.3 -- 10 keV 
flux at t$_{0}$ + 100 s for all the SGRBs in the Swift sample with $T_{90} \le 2 s$ 
and which were observed by XRT at this time. GRB 090515 is shown with a filled circle. 
As expected, the higher fluence GRBs tend to have higher flux X-ray afterglows. 
GRB 090515 is an exception to this alongside GRB 070724A; both of these bursts 
have an unusually high initial X-ray flux for their fluence. In Figure 
\ref{fig4}(a), we compare the combined BAT-XRT light curves of GRB 090515 and 
GRB 070724A. The initial XRT flux of 070724A appears to be consistent with flares 
(as there is a varying hardness ratio) and an underlying broken power law decay. There 
is no obvious plateau phase for GRB 070724A, but this may have occurred prior to the 
XRT observations. The steep decay phase of GRB 070724A, with $\alpha=3.44^{+0.60}_{-0.35}$ 
is much shallower than the steep decay of GRB 090515. Additionally, the optical afterglow 
of GRB 070724A had a magnitude of $i=23.79\pm0.07$ at 2.3 hours after the burst, corresponding to a 
flux of $6.86\times10^{-30}$ erg cm$^{-2}$ s$^{-1}$ Hz$^{-1}$, with an associated host 
galaxy \citep{berger2009, kocevski2009}. This flux is almost an order of magnitude 
larger than the optical afterglow of GRB 090515 at 1.7 hours and GRB 090515 does not have 
an identified host galaxy. However, GRB 070724A does share many similarities with GRB 090515 so 
we cannot rule out the possibility that they originate from a similar progenitor. 

Figure \ref{optical_lcs}(a) shows the lightcurves for the observed R band optical afterglows 
associated with SGRBs (published values converted from magnitudes into flux density 
in Jy), GRB 090515 is the faintest observed and one of the earliest 
detections after the trigger time. In Figure \ref{optical_lcs}(b) we have divided 
the optical fluxes by the XRT flux at 1000 s after the trigger time. When we have considered the 
XRT flux at 1000 s, the optical afterglow of GRB 090515 is not unusually faint compared to other SGRBs.
We also show the optical light curve for GRB 080503 
\citep[a short burst with extended emission][]{perley2009b} in Figure \ref{optical_lcs}(c) in comparison to 
GRB 090515.

\subsection{GRBs with similar fluence to GRB 090515}

As the fluence of GRB 090515 in the 15 -- 150 keV energy band was one of the lowest 
fluences observed for SGRBs, here we compare it to other low fluence GRBs. 

\begin{figure*}
\centering
\includegraphics[width=18cm]{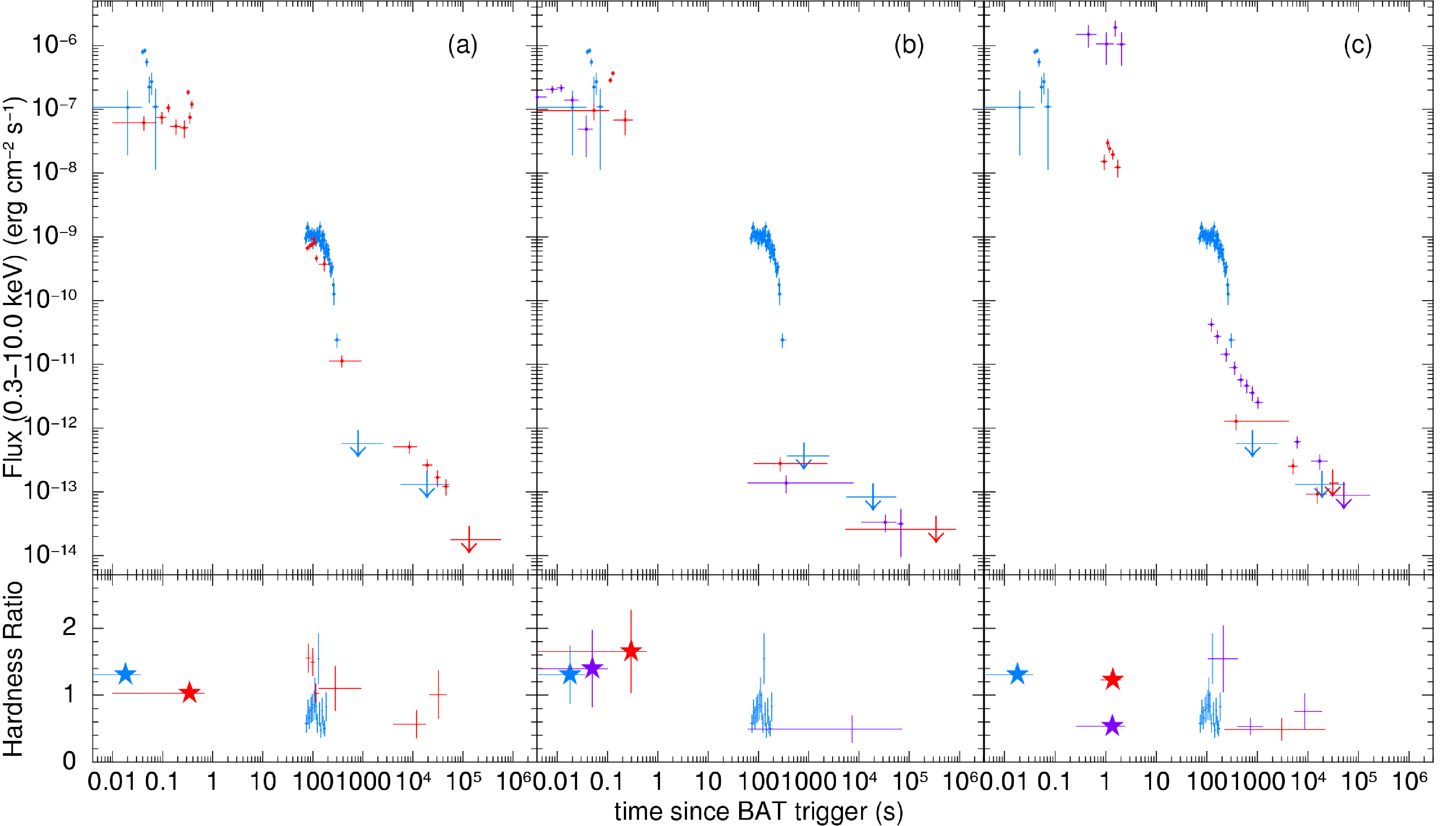}
\includegraphics[width=18cm]{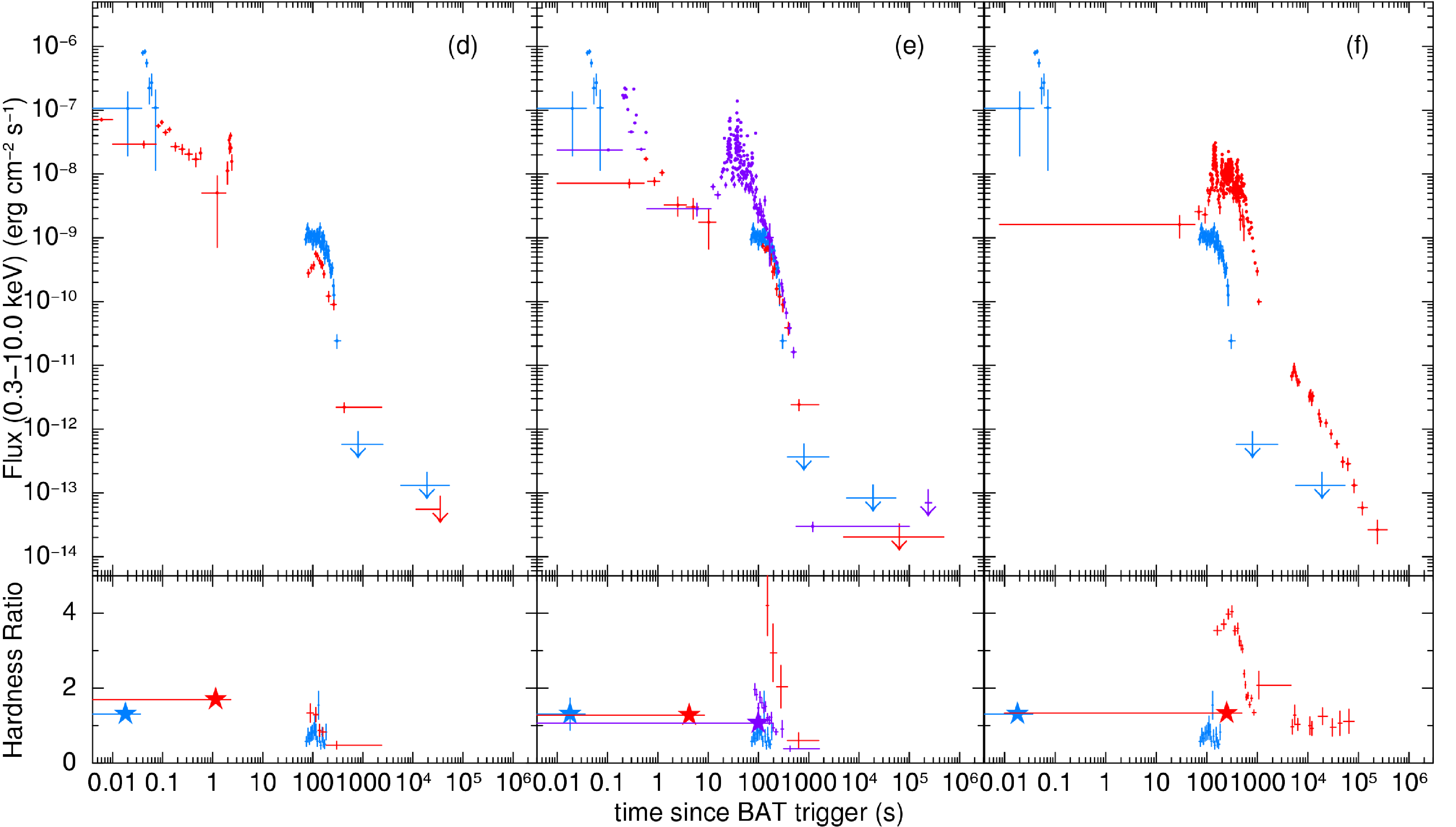}
\caption{The BAT-XRT light curve and hardness ratios for GRB 090515 in blue in comparison to other 
GRBs. (a) GRB 070724A in red. (b) GRB 050813 in red and GRB 050509B in purple. (c) GRB 060717A in red 
and GRB 080520A in purple. (d) GRB 090607 in red. (e) GRB 050421 in red and GRB 080503 in purple. 
(f) GRB 070616 in red. In 
the lower boxes for each graph, there is the hardness ratio for the BAT data 
((50 -- 100) keV/(25 -- 50) keV), with a star, and the hardness ratio for the XRT data 
((1.5 -- 10) keV/(0.3 -- 1.5) keV).}
\label{fig4}
\end{figure*}

GRB 050509B and GRB 050813 were short GRBs detected by the {\it Swift} satellite 
that were similar to GRB 090515 during the prompt emission phase. However, the 
combined BAT and XRT light curves for GRBs 050509B and 050813, shown in Figure 
\ref{fig4}(b), do not show the same X-ray plateau extending to $\sim$200 s after the burst. 
GRBs 050509B and 050813 have both been used to place constraints on the compact binary merger 
model of SGRBs \citep{gehrels2005, hjorth2005b, bloom2006, ferrero2007}. The observed upper limits for
GRB 090515 at late times (after 400s) are consistent with the later emission observed for GRBs 
050509B and 050813. This suggests that the plateau and steep decay are an additional component 
in the light curve of GRB 090515. 

GRB 051105 is a SGRB with an identical fluence to GRB 090515, but its afterglow was undetectable 
by XRT in observations starting 68 s after the burst \citep{mineo2005b}. GRB 070209 had the lowest 
SGRB fluence and was also undetectable by XRT in observations starting 78 s after the burst 
\citep{sato2007a}. 

In Figure \ref{fig4}(c), the X-ray light curve of GRB 090515 is compared to the
two lowest fluence LGRBs in the {\it Swift} sample which were detected by XRT. 
These are GRB 080520A and GRB 060717A, they both have significantly higher fluence 
in the 15 -- 150 keV band than GRB 090515 (due to having longer durations), but are a lot fainter 
in X-rays, again suggesting additional X-ray emission in GRB 090515. 

It is possible that these GRBs had plateau phases which end prior to the XRT observations. However, 
as {\it Swift} slewed promptly to these GRBs (observations typically starting within 100 s), a 
plateau phase would need to be significantly shorter than that observed for GRB 090515. The main 
exception to this is GRB 060717A, which had XRT observations begining when GRB 090515 was in the 
steep decay phase.

\subsection{GRBs with steep decays}

The GRB with an X-ray light curve most similar to GRB 090515 is GRB 090607, which has a 
$T_{90}$ just above the short-long boundary. They are compared in Figure \ref{fig4}(d). 
Both light curves show a distinctive steep decay at $\sim$200s. However, the emission 
of GRB 090607 between 80 and 100 s is not a plateau as observed in GRB 090515 and, given 
the hard spectrum which softens as the emission decays (shown in the lower panel of 
Figure \ref{fig4}(d)), is more likely due to a flare at the start of the XRT 
observations. 

\section{Discussion}

The steep decay in the unusual X-ray light curve of GRB 090515 cannot be explained using 
the external shock afterglow models. Instead, we consider if this GRB was a naked burst 
with faint, rapidly fading emission, and if the X-ray plateau is powered by an unstable 
millisecond pulsar. These possibilities are discussed below.

\subsection{An under-luminous naked LGRB}
If a GRB occurs in a very low density ISM then the afterglow from external 
shocks between the jet and the ISM could be too faint for detection by {\it Swift}. Instead, 
there would just be the prompt emission followed by a rapid decline due to the 
``curvature effect'' \citep{kumar2000}. This predicts a decay in flux described by:
\begin{eqnarray}
f_{\nu}~ \propto~ \nu^{-\beta}t^{-2-\beta} \label{flux}
\end{eqnarray}
where $\beta$ is the observed spectral energy index at frequency $\nu$ ($\beta=\Gamma-1$), 
and t is the time since the trigger. We should observe a decay $\alpha = 2+\beta = \Gamma + 1$. We compare 
here GRB 090515 with a good candidate for a naked burst, GRB 050421 \citep{godet2006}.

GRB 050421 was a weak long GRB detected by BAT following a steep 
decay, as shown in Figure \ref{fig4}(e), although the decay is not as steep as 
for GRB 090515. There is evidence of spectral evolution, as the emission is 
getting softer (the lower panel of Figure \ref{fig4}(e)); however, the spectral evolution is during 
the steep decay and not the plateau region. The initial hardness ratio for (1.5 -- 10) keV/(0.3 -- 1.5) 
keV is 6 times larger for GRB 050421 than GRB 090515.  \cite{godet2006} explained 
the steep decay ($\alpha = 3.1 \pm 0.1$) of GRB 050421 by assuming it was a ``naked burst'', 
i.e. there was no forward shock component of the afterglow as the interstellar medium (ISM) was 
not dense enough for the shock wave to produce a typical afterglow. The detected decaying emission 
is consistent with the ``curvature effect''. GRB 080503, shown in Figure \ref{fig4}(e), has 
also been explained as a short ``naked burst'' with extended emission detected in the BAT (although 
not a plateau), the X-ray decay is consistent with the ``curvature effect'' \citep{perley2009b}. However, 
the steep decay for both of these are significantly shallower than the decay of GRB 090515, which was 
$\alpha_{3} >$9 (with t$_{0}$ at the start of the prompt emission).

GRB 090515 shares some similarities with GRB 050421 and GRB 080503 \citep{godet2006, perley2009b}.
\cite{zhang2009b} suggested that the burst duration, observed by BAT, represents the 
duration that the jet is relativistic and, with a non-relativistic (or less relativistic) 
jet, the central engine can be active for longer than this time and may be observed by XRT. 
Therefore, the X-ray plateau observed for GRB 090515
could be a continuation of the prompt emission, which has fallen below the threshold of 
BAT. So with a more sensitive detector, GRB 090515 may have been identified as a LGRB. If 
true, we should expect to see that the steep decay matches the ``curvature effect'' 
like GRB 050421. During the plateau, the spectral index $\Gamma_{x}$ is $1.88\pm0.14$ predicting 
a steep decay slope of $\alpha = 2.88 \pm 0.14$. As the observed decay is significantly steeper 
than this, it does not fit the ``curvature effect'' theory. Using the method described by 
\cite{liang2006}, we shifted the t$_{0}$ to the possible flare at the end of the plateau in GRB 
090515. The steep decay becomes less extreme, $\alpha = 3.7\pm0.6$, but still steeper than the 
predicted decay slope. This method relies on correctly identifying the time at which 
the central engine is last active and with a plateau in the light curve this point is 
difficult to identify. The steep decay of GRB 090515 following the plateau may be 
consistent with the ``curvature effect'' if a later location of t$_{0}$ is 
identified. Alternatively, this could be associated with a narrow opening angle for the 
jet which creates the plateau, as in that case outside of 1/$\Gamma$ there would be very 
little high latitude emission, giving a much steeper decay slope. It is also possible that the 
spectrum softens immediately prior to the steep decay, however we do not have enough observed counts 
to produce a reliable X-ray spectrum at this time.

GRB 090515 can potentially be explained as an under-luminous naked long GRB, however this is reliant on the 
assumption that the plateau is powered by prolonged activity in the central engine.

\subsection{An unstable millisecond pulsar (magnetar) central engine}
The bright X-ray plateau in the light curve of GRB 090515 could be associated with the 
formation, emission and collapse of a millisecond pulsar. There have been predictions that
in some GRBs an unstable millisecond pulsar may be formed \citep{usov1992, duncan1992, 
dai1998a, dai1998b, zhang2001}. At formation, there is enough 
rotational energy to prevent gravitational collapse. This energy can be released as 
electromagnetic radiation or gravitational waves, causing the pulsar to spin down until it 
reaches a critical point at which it is no longer able to support itself. At this 
point the pulsar collapses to a black hole and the emission stops. This would be 
evident in the X-ray light curve as a plateau caused by energy injection from the
millisecond pulsar followed by an extremely steep decay when the pulsar collapses. 
We might expect millisecond pulsars formed during the core collapse 
of a massive progenitor star to be associated with long GRBs and this has been suggested
by \cite{troja2007} and \cite{lyons2009}. GRB 090515 was an extremely short GRB, but 
a millisecond pulsar could be formed by two merging neutron stars 
(a potential progenitor of SGRBs), depending on various assumptions about the 
neutron stars' equations of state \citep{dai1998a, dai2006, yu2007}.

\cite{troja2007} and \cite{lyons2009} studied LGRBs with a plateau and a steep decay and GRB 
090515 shows similarities to them. In Figure \ref{fig4}(f), we compare the light curve of GRB 
090515 to that of GRB 070616 \citep{starling2008}, one of the sample chosen by \cite{lyons2009} 
as potentially showing evidence of an unstable millisecond pulsar. When comparing the 
light curves, GRB 070616 appears to be a brighter and longer version of GRB 090515 but with
a bright afterglow component at later times. 

We have used the following equations from \cite{zhang2001} (equations \ref{period} and 
\ref{luminosity}) to determine if GRB 090515 could be a millisecond pulsar, using 
$T_{em,3}$, the rest frame duration of the plateau in units of $10^{3}$ s, and $L_{em,49}$, the luminosity of 
the plateau in units of $10^{49}$ erg s$^{-1}$, in the rest frame energy band 
1 -- 1000keV. The equations are rearranged to 
give equations \ref{b^2} and \ref{p^2}, these are used to predict the magnetic field 
strength and the spin period of a pulsar formed by this method.
\begin{eqnarray}
T_{em,3}=2.05~(I_{45}B^{-2}_{p,15}P^2_{0,-3}R^{-6}_6)\label{period}\\
L_{em,49}\sim(B^2_{p,15}P^{-4}_{0,-3}R^6_6)\label{luminosity}\\
B^{2}_{p,15}=4.2025 I_{45}^{2}R^{-6}_{6}L_{em,49}^{-1}T_{em,3}^{-2}\label{b^2}\\
P^{2}_{0,-3}=2.05 I_{45}L_{em,49}^{-1}T_{em,3}^{-1}\label{p^2}
\end{eqnarray}
where $I_{45}$ is the moment of inertia in units of $10^{45}$g cm$^{2}$, $B_{p, 15}$ 
is the magnetic field strength at the poles in units of $10^{15} G$, $R_{6}$ is 
the radius of the neutron star in $10^{6}$cm and $P_{0,-3}$ is the initial 
period of the compact object in milliseconds. These equations apply to the 
electromagnetic dominated spin down regime, as the gravitational wave dominated regime 
would be extremely rapid and produce a negligble effect in our analysis. We could assume, 
as in \cite{lyons2009}, that we can use standard values for a neutron star so that  
$I_{45} \sim 1$ and  $R_{6} \sim 1$ which may be appropriate for a collapsar. However, as 
we would be forming an unstable millisecond pulsar by merging two neutron stars the true 
values may be different, depending on the mass and equation of state. For a millisecond 
pulsar formed by a binary merger, we can take the mass of the neutron star to be 
M$_{NS}=$2.1M$_{\odot}$ \citep{nice2005} and estimate $I_{45} \sim 1.5$. Although GRB 
090515 has many properties similar to other SGRBs suggesting the progenitor is most likely 
a compact binary merger, there have been predictions that collapsars may also produce a 
SGRB \citep[for example from an orphan precursor jet, ][]{janiuk2008} and evidence that a 
significant fraction of SGRBs are related to collapsars rather than compact binary mergers 
\citep{virgili2009, cui2010}. So in the following analysis we compare both progenitor models.

\begin{figure}
\centering
\includegraphics[width=7.5cm]{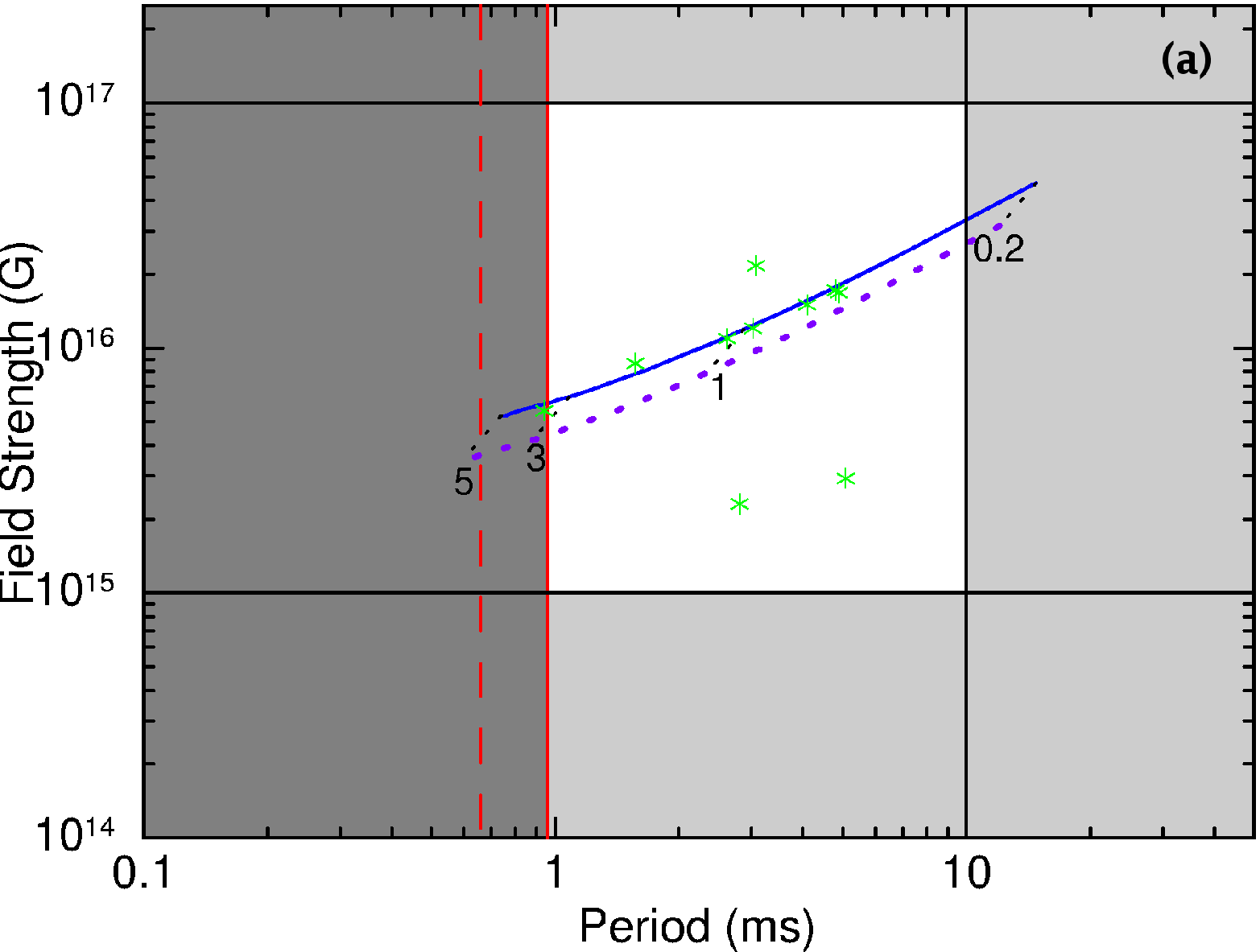}
\includegraphics[width=7.5cm]{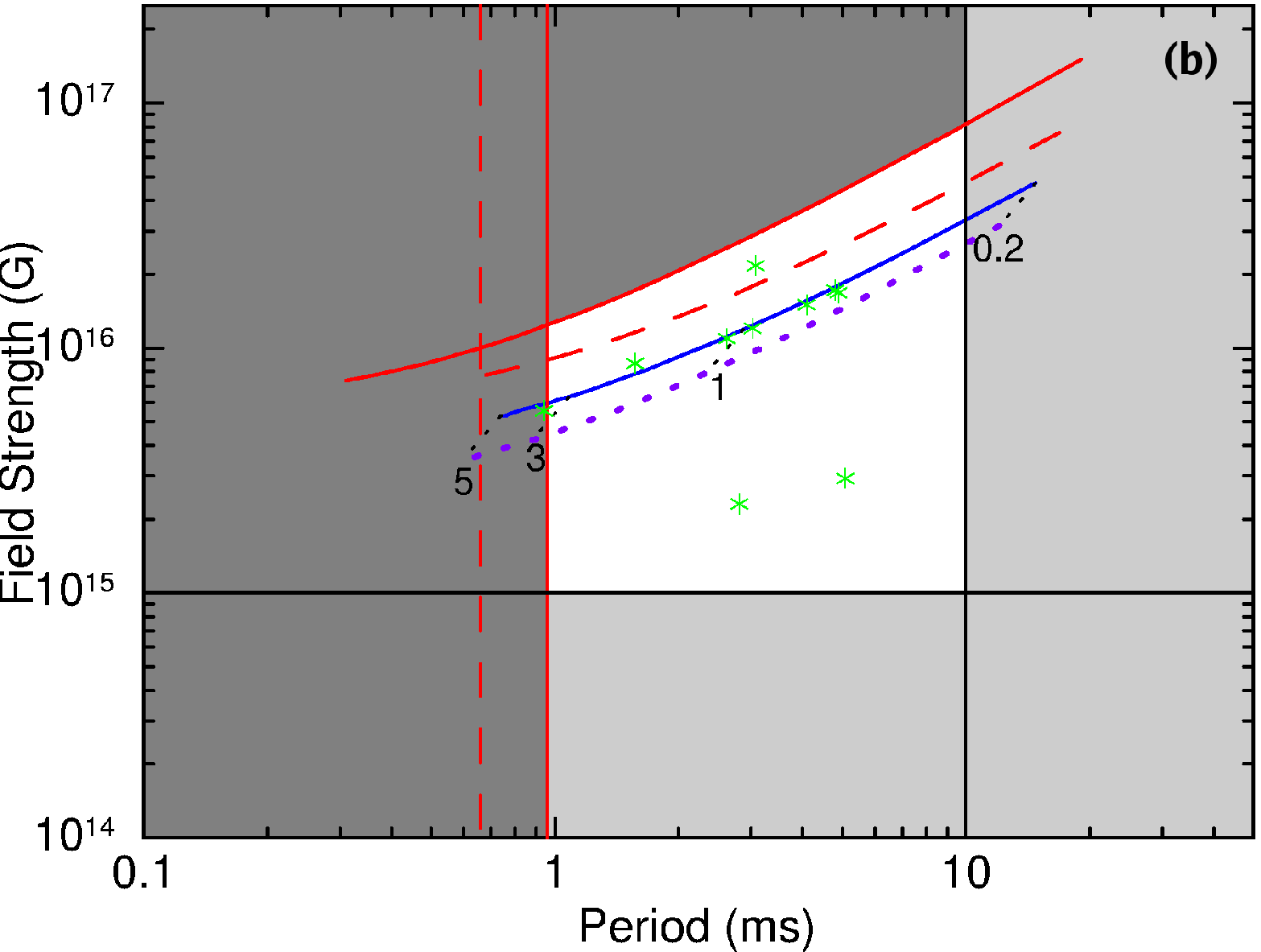}
\includegraphics[width=7.5cm]{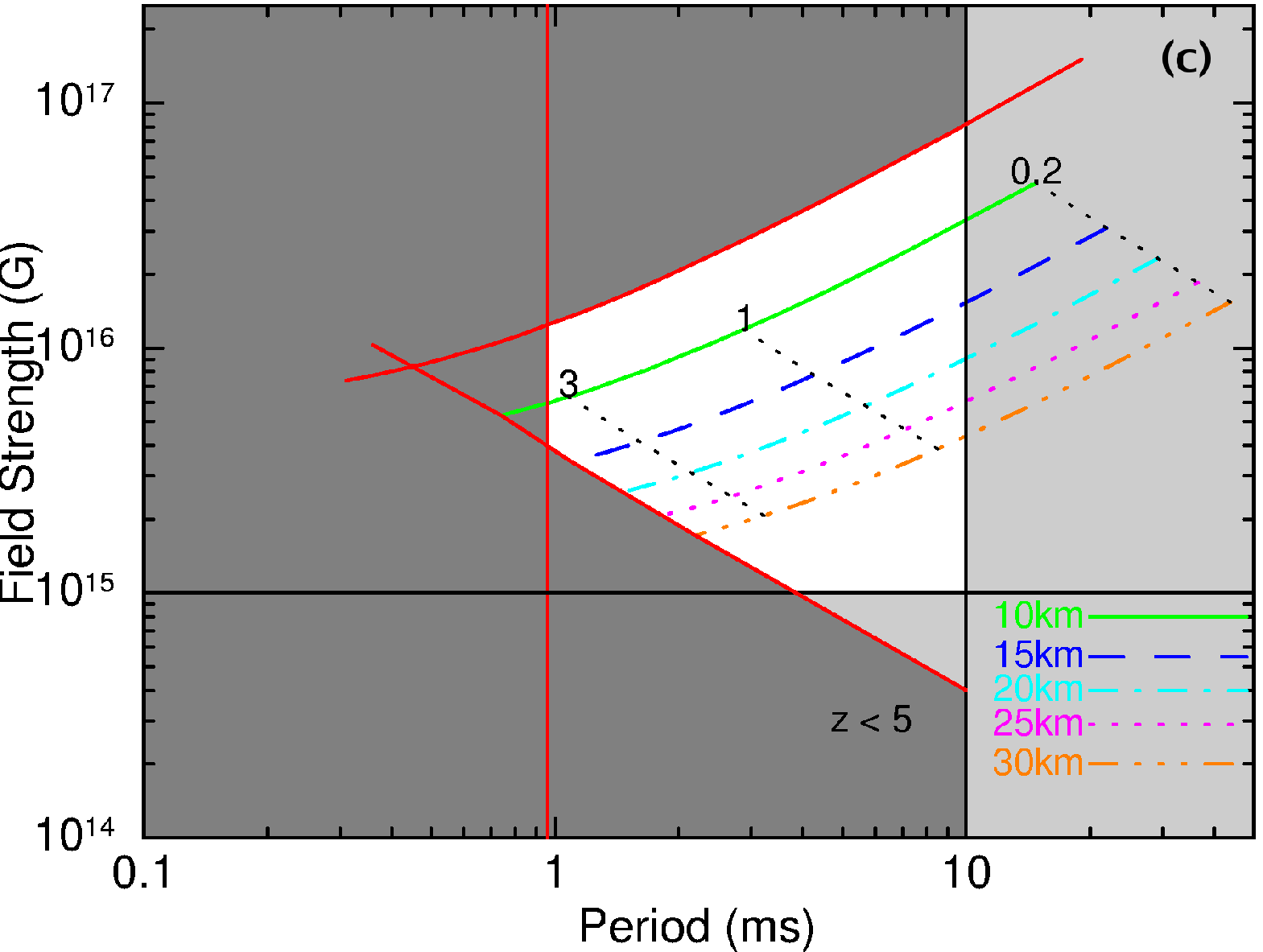}
\caption{(a) The blue line shows the magnetic field and period for a millisecond pulsar 
formed during GRB 090515 as a function of redshift assuming a neutron star mass of 
1.4M$_{\odot}$ and the purple dotted line assumes a neutron star mass of 2.1M$_{\odot}$. The green 
stars are the 18 degree beamed LGRB sample from \citet{lyons2009}. The red line shows the 
limit at which the progenitor would violate the breakup spin period of a 
pulsar for a mass of 1.4$M_{\odot}$ and the dashed red line is for a mass of 2.1$M_{\odot}$. 
The other regions are as defined in \citet{lyons2009}; dark grey shading corresponds to forbidden 
regions (assuming a mass of 1.4$M_{\odot}$) and light grey are limits based on the previous 
studies (as discussed in the text). The dotted lines represent contours of equal redshift 
decreasing from left to right.
(b) The upper magnetic field limit in (a) has been replaced by the red curved line giving the 
forbidden region assuming causality shaded in dark grey (assuming a mass of 1.4$M_{\odot}$). This 
region will change depending on the mass of the neutron star, the rest frame duration 
and luminosity of the plateau. The red dashed curved line represents the forbidden region for a 
binary merger progenitor.
(c) The different contours represent the effect of increasing the radius of the neutron 
star from 10 km to 30 km assuming a constant mass of 1.4$M_{\odot}$. Additionally, we include 
a limit imposed on redshift due to detection of the afterglow in the R-band.}
\label{fig16}
\end{figure}

As a redshift was not obtained for this GRB, we used a range of redshifts from $z = 0.2$ up 
to an upper limit of $z = 5.0$ consistent with the detection of the optical afterglow. We assume 
that the millisecond pulsar was formed at $t \sim 0$ 
and, hence, the duration of the plateau in the observer frame is 240 s. We calculate 
the luminosity of the plateau using the observed 0.3 -- 10 keV flux of 
$\sim 1 \times 10^{-9}$ erg cm$^{-2}$ s$^{-1}$, the spectral index during the 
plateau (1.88) and a k-correction \citep{bloom2001}. These values were then substituted 
into the equations (\ref{b^2}) and (\ref{p^2}) to calculate $B_{p, 15}$ and $P_{0,-3}$. 
These are plotted as a blue contour in Figure \ref{fig16}(a) assuming it was formed 
from a collapsar and a purple contour if formed by a binary neutron star merger. 

Also shown in Figure \ref{fig16}(a), are the regions in which a millisecond pulsar would be 
expected, as defined in 
\cite{lyons2009}: the red line represents the breakup spin-period for a neutron star of mass 
1.4$M_{\odot}$ \citep[$\ge$0.96ms, ][]{lattimer2004}. Using equation \ref{lattimer} 
\citep{lattimer2004}, we calculate this limit for the binary merger scenario with a 
mass of 2.1$M_{\odot}$ to be P$\ge$0.66 ms (where P is the minimum spin period of the 
neutron star in ms) and this is shown with a red dashed line.
\begin{eqnarray}
P_{0,-3} \ge 0.81 M_{1.4}^{-1/2} R_{6}^{3/2} ms\label{lattimer}
\end{eqnarray}
The initial rotation period needs to be $\le$10ms \citep{usov1992} and the lower limit for the magnetic 
field is $\ge$10$^{15}$G \citep{thompson2007}. This shows that GRB 090515 could have formed 
a millisecond pulsar if it had a redshift of $0.3<z<3.5$ for a collapsar progenitor or a
redshift of $0.2<z<4.4$ for a binary merger progenitor. These are both very reasonable 
redshift ranges when we compare them to the sample of GRBs. The magnetic field for 
a given spin period is slightly lower for a binary merger progenitor than for a collapsar 
progenitor. Alongside the prediction by \cite{troja2007} and \cite{lyons2009} of a plateau 
followed by a steep decay for the lightcurve of a millisecond pulsar collapsing to a black 
hole, which matches the observed light curve for GRB 090515, this analysis provides a 
consistent case for GRB 090515 forming a millisecond pulsar irrespective of the two initial 
progenitor models considered.

Using a causality argument, i.e. that the speed of sound on the neutron star cannot exceed 
the speed of light, we can place a tighter constraint on the minimum possible radius, $R_{6}$ 
and  $M_{1.4}$, where $M_{1.4}$ is the mass of the neutron star in 1.4$M_{\odot}$, using
equation \ref{causality2} \citep{lattimer1990}. The moment of inertia, given in equation 
\ref{inertia}, is based on the assumption that the neutron star can be modelled as an uniform sphere.

\begin{eqnarray}
R_{6}>0.6225M_{1.4} \label{causality2}\\
I_{45} \sim M_{1.4}R_{6}^2 \label{inertia}
\end{eqnarray}

This constraint on radius and moment of inertia for a given mass can be substituted 
into equations \ref{b^2} and \ref{p^2} to define a forbidden region for a given neutron star mass, plateau 
duration and luminosity. The forbidden region is described by equations \ref{b2} and \ref{p2} 
and is shown in Figure \ref{fig16}(b) for GRB 090515 assuming a mass of $1.4M_{\odot}$, for a 
collapsar progenitor (red curved line), and $2.1M_{\odot}$, for a 
binary merger progenitor (red curved dashed line).

\begin{eqnarray}
B^{2}_{p,15} > 10.8 T_{em,3}^{-2} L_{em,49}^{-1} \label{b2}\\
P^{2}_{0,-3} < 0.794 M_{1.4}^{3} T_{em,3}^{-1} L_{em,49}^{-1} \label{p2}
\label{p2}
\end{eqnarray}

It has been suggested that the radii of proto neutron stars may be as large as 
a few tens of kilometers \citep{ott2006}, so in Figure 
\ref{fig16}(c) we show the effect of increasing the radius, from 10 km to 30 km, for 
a mass of 1.4$M_{\odot}$, using the plateau luminosities and durations previously calculated 
for GRB 090515 assuming it is at a range of redshifts. For 
larger radii, the unstable millisecond pulsar has to be at higher redshifts, have a smaller 
magnetic field and larger period. As we have an R-band detection of the optical afterglow, 
we can place the upper-limit z$\le$5 on the redshift .

\begin{figure}
\centering
\includegraphics[width=8.4cm]{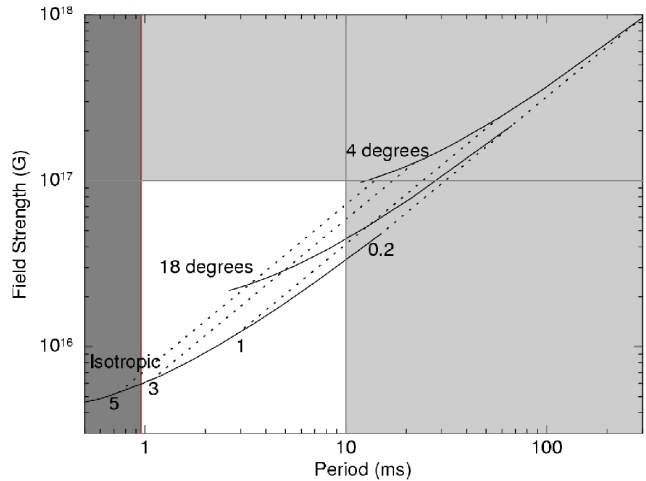}
\caption{We show here the effect of beaming the emission of GRB 090515 assuming a mass of 
1.4$M_{\odot}$. Solid lines show isotropic solution and the solutions for the two beaming 
angles considered in \citet{lyons2009}. The dotted lines represent contours of equal 
redshift decreasing from left to right. The forbidden regions are as defined for Figure 
\ref{fig16}(a).}
\label{fig16b}
\end{figure}

In Figure \ref{fig16b}, we investigate the effect of the different beaming angles considered by 
\cite{lyons2009} assuming a mass of 1.4$M_{\odot}$. As the causality forbidden region shown in 
Figure \ref{fig16}(b and c) also depends on beaming angle we have reverted to using the regions 
defined by \cite{lyons2009} for clarity. Up to this point, we have only considered isotropic 
emission and this shows beaming the emission would greatly affect the results obtained. 
Simulations have shown that a relativistic jet can be produced by a magnetar \citep{bucciantini2009}. 
If the emission was beamed by 4 degrees the observations would not support the 
magnetar model. With a beaming angle of 18 degrees, GRB 090515 would need a redshift of 
$1<z<5$ in order to satisfy the model and the constraints obtained by observing an optical 
afterglow. The more tightly the emission is beamed, the higher the redshift that the burst 
would need to be at in order to fit the magnetar model and this may explain why a host 
galaxy has not been identified. 

\section{Conclusions}

GRB 090515 is a very unusual SGRB, as its low gamma-ray fluence would lead us to expect a 
significantly fainter X-ray light curve than observed at early times. Most importantly, 
the X-ray plateau followed by an extremely steep decay is very unusual, but may not 
be unique in the {\it Swift} sample. With a more sensitive detector, 
the plateau observed by XRT may have instead been identified 
as part of the prompt emission and GRB 090515 might instead have been classified as a 
LGRB. Therefore, it poses interesting questions about the progenitor model and for the 
classification of other GRBs. In this paper, we have considered the two popular 
progenitor models for GRBs, collapsars and compact binary mergers.

GRB 090515 is the first SGRB with a fluence below $10^{-7}$ erg cm$^{-2}$ with an observed 
optical afterglow at 1.75 hours (R=26.4$\pm$0.1), and  this is the faintest detected optical 
afterglow for a GRB at that time.

We suggest that the simplest explanation for the unusual light curve of 
GRB 090515 is that it shows prolonged emission from an 
unstable millisecond pulsar, followed by an extremely steep decay when the millisecond
pulsar collapses. Given the short duration of the GRB and the other properties, we favour 
the binary merger progenitor but cannot rule out a collapsar progenitor. For a collapsar 
progenitor, the proposed unstable millisecond pulsar with a spin period of 10 ms would have 
a magnetic field of $\sim 3\times10^{16}$ G at z $\sim$ 0.3 and with a spin period of 1 ms the 
magnetic field would be $\sim 6\times10^{15}$ G at z $\sim$ 3.5. The binary merger progenitor 
model gives a spin period of 10 ms and a magnetic field of $\sim 2.5\times10^{16}$ G at 
z $\sim$ 0.2 to a spin period of 66 ms and a magnetic field of $\sim 4\times10^{15}$ G at 
z $\sim$  4.4. These values assume isotropic emission and a radius of 10 km.

\section{Acknowledgements}

AR, NRT, PAE, NL, AJL, KLP and APB would like to acknowledge funding from the Science and Technology 
Funding Council. This work makes use of data supplied by the UK {\it Swift} Science Data 
Centre at the University of Leicester and the {\it Swift} satellite funded by NASA 
and the Science and Technology Funding Council. {\it Swift} funding at PSU comes from 
NASA contract NAS5-00136. The Dark Cosmology Centre is funded by the DNRF

Based on observations obtained at the Gemini Observatory, which is operated by the
Association of Universities for Research in Astronomy, Inc., under a cooperative agreement
with the NSF on behalf of the Gemini partnership: the National Science Foundation (United
States), the Science and Technology Facilities Council (United Kingdom), the
National Research Council (Canada), CONICYT (Chile), the Australian Research Council
(Australia), Ministério da Ciência e Tecnologia (Brazil) 
and Ministerio de Ciencia, Tecnología e Innovación Productiva  (Argentina)

\end{document}